\documentclass[11pt,a4paper]{article}
\usepackage[latin1]{inputenc}
\usepackage{amsmath}
\usepackage{amsfonts}
\usepackage{amssymb}
\usepackage{graphicx}
\usepackage{hyperref}
\usepackage{listings}
\usepackage{tikz}

\lstdefinestyle{customc}{
  belowcaptionskip=1\baselineskip,
  breaklines=true,
  frame=L,
  xleftmargin=\parindent,
  language=C,
  showstringspaces=false,
  basicstyle=\footnotesize\ttfamily,
  keywordstyle=\bfseries\color{green!40!black},
  commentstyle=\itshape\color{purple!40!black},
  identifierstyle=\color{blue},
  stringstyle=\color{orange},
}

\lstdefinestyle{customasm}{
  belowcaptionskip=1\baselineskip,
  frame=L,
  xleftmargin=\parindent,
  language=[x86masm]Assembler,
  basicstyle=\footnotesize\ttfamily,
  commentstyle=\itshape\color{purple!40!black},
}

\newcommand*\Laplace{\mathop{}\!\mathbin\bigtriangleup}

\graphicspath{{./brushimages/}}
\usepackage[left=2cm,right=2cm,top=2cm,bottom=2cm]{geometry}
\author{Luke Kristopher Davis}
\title{1D Classical Density Functional Theory of a Tethered Polymer Layer}

\begin{document}
\lstset{language=C++}
\maketitle
\newpage
\tableofcontents
\newpage
\section{Introduction}

The system under consideration comprises of polymers, with uniform monomeric structure, permanently attached to points ${r_\bot= (x_\bot, y_\bot)}$ on a flat 2-dimensional surface. In the literature this system is commonly referred to as a \textit{'tethered polymer layer'} or simply a \textit{'polymer brush'}, however as we shall see a brush is only formed when a certain condition is met.

Polymer layers instil interest in scientists from fields ranging from engineering, chemistry and biology. It is possible to modify the properties of interfaces using polymers, say, by reducing the surface energy of a polymer melt by blending in a polymer of lower surface energy that segregates to the surface \cite{Jones1999}. The FG Nups (long chained proteins) inside the nuclear pore complex, a biogical machine which acts as a selective transport barrier to the nucleus of a cell, can be considered as polymers grafted to the inside of the cylindrical pore. The problem of understanding how this transport barrier arises is then reduced to a problem of how the polymer layer dynamically interacts with transport molecules \cite{Osmanovic2013}.

The field of tethered polymer layers is over 2 decades old and initial scaling theories were provided by \cite{Alexander1977} \cite{DeGennes1980}. Alexander and De Gennes found that the density of the polymer layer, $\rho$, to the normal of the plane was found to have a step-function form. An analytical Self consistent field theory (SCFT) study by Milner, Witten and Cates \cite{Milner1988} consisted of formulating the  partition function $Z_{sc}$ for a single chain in the presence of a mean field $w(r)$ due to the other polymers. The form of this field is not known a priori and as such must be found self-consistently. The partition function $Z_{sc}$ is:

\begin{equation}
Z_{sc} = \Sigma_{{\textbf{r}(s)}} exp(-S_{k})
\end{equation}

where $\textbf{r}(s)$ is the position along the chain at contour label $s$. $S_{k}$ for a configuration $k$ is given by:

\begin{equation}
\label{eq_action}
S_k = \int [ \frac{1}{2} (\frac{d \textbf{r}_{k}(s)}{ds})^2  - w(\textbf{r}(s))]
\end{equation}

which is analogous to the action in quantum mechanics. The partition function, under conditions of strong stretching, is then dominated by configuration(s) which minimize the action. The first term in the integrand of equation \ref{eq_action} is the stretching energy which is analogous to the kinetic energy of a particle. So a strongly stretched configuration can be thought of as a particle with a large momentum and therefore can be handled within the classical limit of quantum mechanics. 

\begin{figure}[h]
{
\begin{center}
\includegraphics[scale=0.4]{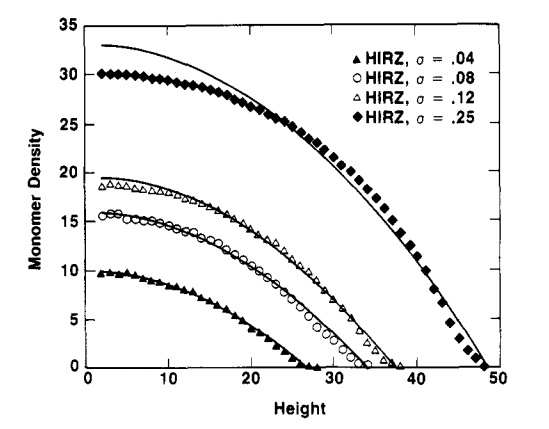}
\caption{ Segment density profiles for grafted chains in a good solvent calculated by Hirz using numerical SCFT. Solid lines are the analytical calculations from MWC \cite{Milner1988}. The key relates the plots to specific grafting densities $\sigma$. \label{fig_MWC}}
\end{center}
}
\end{figure}

Within this regime MWC computed segment density profiles (figure \ref{fig_MWC} ) which had a parabolic form, conflicting the step function prediction given by Alexander and De Gennes.

\newpage

The aim here is to outline the implementation of a classical density functional theory (cDFT) scheme to investigate the behaviour of a tethered polymer layer. The theory, computational scheme and results are presented.

\section{Polymer theory}
\subsection{Polymer chains in external fields}

Let the potential energy contribution $U$ due an external field $w$ be given by:

\begin{equation}
U(\lbrace \textbf{r} \rbrace) = k_{B} T \sum_{i=0}^{N} w(\textbf{r}_i)
\label{eq_PE}
\end{equation}

where $\lbrace \textbf{r} \rbrace$ means the set of all positions of $N+1$ monomers, $k_B$ is Boltzmann's constant and $T$ is the temperature. The sum $\sum_{i=0}^{N}$ runs from the first monomer to the last.

In the continuous case where the polymer chain is comprised of infinitely small segments the sum $\sum_{i=0}^{N}$ becomes $\int_0^N ds$. The equilibrium distribution of an ideal chain is given by:

\begin{equation}
\Psi [\textbf{r(s)}] = A \cdot exp(-\frac{3}{2b^2} \int_0^N (\frac{\partial \textbf{r(s)}}{\partial s})^2 ds)
\label{eq_weiner}
\end{equation}

where $A$ is a constant and $b$ is the statistical segment length defined by $\frac{R_e}{\sqrt{N}}$ where $R_e$ is the end-to-end distance of the polymer \cite{Doi1986}. Equation \ref{eq_weiner} is also known as the Weiner distribution. Under an external field it is modified by a Boltzmann factor such that the new conformational distribution function is:

\begin{equation}
\Phi [\textbf{r(s)}] = A \cdot exp(-\frac{3}{2b^2} \int_0^N (\frac{\partial \textbf{r(s)}}{\partial s})^2 ds - \beta \int_0^N w(\textbf{r(s)}) ds)
\label{eq_conf}
\end{equation}

Let $G(\textbf{r}, \textbf{r}^\prime, N)$ be a green's function which represents the likelihood (propagator) that a chain of $N$ steps which started at position \textbf{r} ends at position $\textbf{r}^\prime$. This green's function is defined as:

\begin{equation}
G(\textbf{r},\textbf{r}^\prime,N) = \frac{\int_{\textbf{r}_0 = \textbf{r}}^{\textbf{r}_{N} = \textbf{r}^\prime} \Phi [\textbf{r(s)}] \mathcal{D}[\textbf{r}]}{\int d\textbf{r}^\prime \int_{\textbf{r}_0 = \textbf{r}}^{\textbf{r}_{N} = \textbf{r}^\prime} \Psi [\textbf{r(s)}] \mathcal{D}[\textbf{r}]}
\label{eq_G}
\end{equation}

note the functional measures which denote functional integration since \textbf{r} are explicit functions of the contour label $s$. Therefore the sum over all possible configurations is given by $Z$, the partition function, which is computed by summing over all starting and ending positions for $N+1$ steps as shown in equation \ref{eq_ZG}.

\begin{equation}
Z = \int d\textbf{r} \int d\textbf{r}^\prime G(\textbf{r}, \textbf{r}^\prime, N)
\label{eq_ZG}
\end{equation}

\begin{figure}[h]
\begin{center}
\includegraphics[scale=0.35]{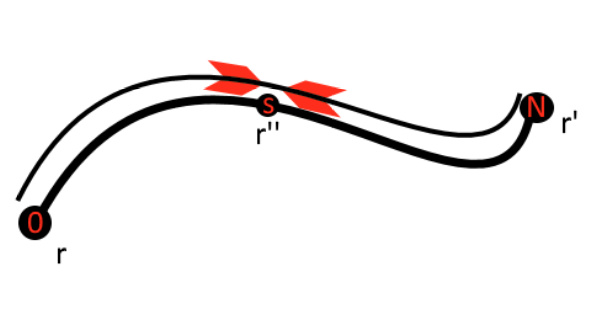}
\caption{ The propagator $G(\textbf{r},\textbf{r}^{\prime \prime},s)$ starts at \textbf{r} for s=0 and ends up at    $\textbf{r}^{\prime \prime}$. The same logic applies to $G(\textbf{r}^{\prime},\textbf{r}^{\prime \prime},N-s)$ but it comes in from the end of the chain. \label{fig_propa}}
\end{center}
\end{figure}

One of the properties of equation \ref{eq_G} is that any statistical weight can be 'built' from a convolution of two statistical weights:

\begin{equation}
G(\textbf{r},\textbf{r}^\prime,N) = \int d\textbf{r}^{\prime \prime} G(\textbf{r},\textbf{r}^{\prime \prime},s) \cdot G(\textbf{r}^{\prime},\textbf{r}^{\prime \prime},N-s)
\end{equation}

where one propagator 'picks' up from where the other left off. This property, which is visualised in figure \ref{fig_propa}, becomes useful when we start computing observables later.

\newpage

Modifying the notation for clarity: $G(\textbf{r}, N)$ is the propagator for a chain of $N+1$ beads to have its end bead at position \textbf{r}. The propagators can be built up recursively via:

\begin{equation}
G(\textbf{r}, s + \Delta s) = \int d\textbf{r}^\prime \Phi[\textbf{r} - \textbf{r}^\prime] G(\textbf{r}^\prime, s)
\label{eq_CK}
\end{equation}

where $\Phi[ \cdots ]$ is given by equation \ref{eq_conf}. Equation \ref{eq_CK} is a Chapman- Kolmogorov equation \cite{Fredrickson2007}  which gives the transitional densities of a Markov sequence, since building the chain as a random walk in the presence of an external disturbance is assumed Markovian. 

Taylor expanding the LHS for small $\Delta s$ to $1^{st}$ order and the RHS for small $\Delta \textbf{r} = \textbf{r} - \textbf{r}^\prime$ to $2^{nd}$ order leads to the \textit{forward-Chapman-Kolmogorov} equation:

\begin{equation}
\boxed{\partial_s G(\textbf{r},s) = \frac{b^2}{6} \Laplace G(\textbf{r},s) - w(\textbf{r})G(\textbf{r},s) }
\label{eq_FP}
\end{equation}

where $\Laplace = (\partial_x^2, \partial_y^2, \partial_z^2)$ and $w(\textbf{r})$ is in units of $\frac{1}{k_B T}$. This is analogous to the 'Feynman- Kac' formula in QM and the modified diffusion equation. The only conceptual difference is that time here is in terms of the contour label $s$. The initial condition is $G(\textbf{r}, s=0) = 1$ which means that the probability to go from e.g. \textbf{r} to $\textbf{r}^\prime$ in 0 steps is 1, since trivially $\textbf{r} = \textbf{r}^\prime$.

When there is no external field present the polymer chain is just a random walk whose green's function obeys the diffusion equation:

\begin{equation}
\boxed{\partial_s G(\textbf{r},s) = \frac{b^2}{6} \Laplace G(\textbf{r},s) }
\label{eq_FPideal}
\end{equation}
\newpage
\section{Tethered Polymer System}

\begin{figure}[h]
\begin{center}
\includegraphics[scale=0.4]{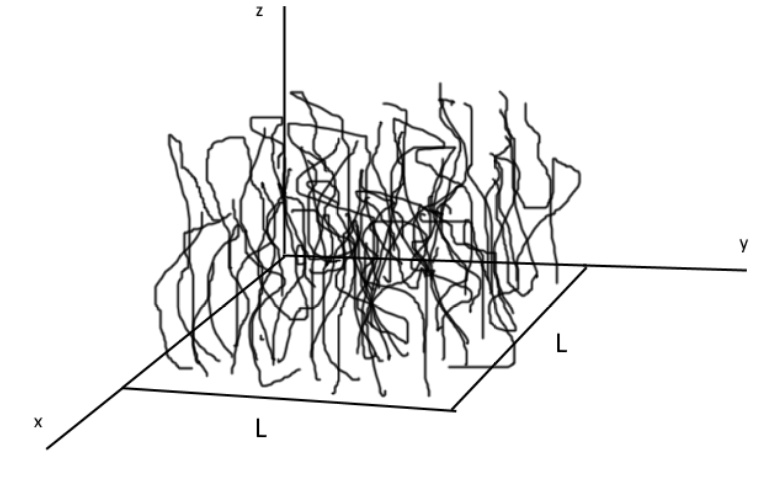}
\caption{ Illustration of polymers grafted on plane for $z=0$.\label{fig_system}}
\end{center}
\end{figure}

For the system under investigation it is assumed polymers are grafted to a square plane with side length $L_{\bot}$ (figure \ref{fig_system}). The length of the system perpendicular to the plane is $L_{z}$ which gives a system volume $V= L_z L_{\bot}^2$. It is forbidden for polymers to penetrate the surface and the former are assumed to be surrounded by implicit solvent, the polymers have no interaction with the surface apart from the impenetrability condition. Polymers have a segment length $b$ of unity and the grafting density of polymers $\sigma$ is defined as the ratio:

\begin{equation}
\sigma= \frac{N_p b^2}{L_\bot^2}
\end{equation}

which assumes uniform grafting within the system. The chains are mono-disperse with $N$ monomers per polymer.

\begin{figure}[h]
\begin{center}
\includegraphics[scale=0.4]{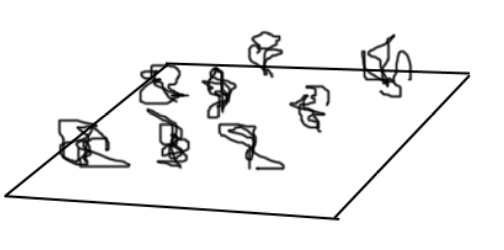}
\caption{ The individual polymers behave as if no other polymers were present for $D>R_{F}$. \label{fig_mush}}
\end{center}
\end{figure}

An important distance in the problem is the distance between grafting points $D=\frac{1}{\sqrt{\sigma}}$. If this distance is less than the Flory radius $R_{F} \approx N^{(\frac{3}{5})}$ it will lead to interactions between the polymers and hence a polymer brush will form. If $D> R_{F}$ the polymers will form small mushroom like configurations (figure \ref{fig_mush} \cite{Jones1999}.

\newpage

The normalized partition function for a chain grafted at $\textbf{r}_{\bot}$ can be computed by means of

\begin{equation}
Q(\textbf{r}_{\bot}) = \int G({\textbf{r}_{\bot}},\textbf{r},N) d\textbf{r}
\end{equation}

where $G({\textbf{r}_{\bot}},\textbf{r},N)$ is a propagator for a chain of length N that started at $\textbf{r}_\bot$ in the grafting plane and ends at \textbf{r}. This $G_{\textbf{r}_\bot}$ satisfies equation \ref{eq_FP} but with initial condition (IC):

\begin{equation}
G({\textbf{r}_{\bot}},\textbf{r},s=0) = \delta(x - x_\bot) \delta(y - y_\bot) \delta(z - 0)
\end{equation}

 and since the units of a Dirac-delta function are the units of inverse its argument it is easy to see that $G_{\textbf{r}_\bot}$ has units $V^{-1}$. The impenetrability of the grafting surface is imposed as Dirichlet boundary conditions on equation \ref{eq_FP} i.e. $G({\textbf{r}_{\bot}},\textbf{r=r}_{\bot},s)=0$. There seems to be a direct contradiction between our initial condition and our boundary condition, to avoid this the polymers are attached a small distance above the plane at $z=0$ so that $z_{\bot} > 0$. So our conditions on the modified diffusion equation are thus:

\begin{description}
\item[IC:] $G({\textbf{r}_{\bot}},\textbf{r},s=0) = \delta(x - x_\bot) \delta(y - y_\bot) \delta(z - z_\bot)$

\item[BC:] $G({\textbf{r}_{\bot}},(x=x_\bot,y=y_\bot, z=0),s)=0$
\end{description}

The segment density $\rho$ is obtained by functional differentiation of $ln[Q]$ w.r.t the external field $w$ \footnote{It is useful to note that $Q$, $G$ and $\rho$ are all functionals of the external field $w$. Normally this is written as $G(\textbf{r}_\bot, \textbf{r}, N; [w])$ for example but the underlying field dependence isn't explicitly stated here for clarity.}:

\begin{equation}
\rho(\textbf{r}) = - \sigma \int d\textbf{r}_\bot \frac{\delta ln [Q(\textbf{r}_\bot)]}{\delta w(\textbf{r})}
= \int d\textbf{r}_\bot \frac{\sigma}{Q(\textbf{r}_\bot)} \int_0^N ds G({\textbf{r}_{\bot}},\textbf{r},s) G({\textbf{r}^{\prime}},\textbf{r},N-s)
\label{eq_density1}
\end{equation}

where it is clear that the segment density and external field are thermodynamically conjugate variables. $G({\textbf{r}^{\prime}},\textbf{r},s)$ is a complementary propagator that starts from the free end of the chain at $\textbf{r}^\prime$ as illustrated in figure \ref{fig_propa} and is found by solving equation \ref{eq_FP} with \textbf{IC :} $G( \cdots, s=0) =1$. 

A more elegant and computationally efficient way to compute the segment density was proposed by Muller \cite{Muller2002}:

\begin{equation}
\rho(\textbf{r}) = \int_0^N ds G_{C}(\textbf{r},s) G(\textbf{r}, N-s)
\label{eq_density_muller}
\end{equation}

where $G_C$ is a propagator with \textbf{IC :} 
\begin{equation}
G_C(\textbf{r},s=0)=\frac{\sigma \delta(z-z_\bot)}{G((x,y,z_\bot),N)}
\label{eq_i2}
\end{equation}

which implies that $G(\textbf{r},s)$ is solved first with unity initial conditions and the value of this green's function in the tethering plane is inserted into equation \ref{eq_i2}.

To simplify matters further, the system is laterally symmetric and therefore $\rho(\textbf{r}) \rightarrow \rho(z)$ and $w(\textbf{r}) \rightarrow w(z)$ which means the greens functions themselves are essentially one dimensional. This is made explicit within the machinery since the function $ G((x,y,z_\bot),N)$ is really just a fixed value $G(z_\bot,N)$.
\newpage
\section{Numerical Solution of The Diffusion Equation}

To numerically solve equation \ref{eq_FP}, which is a parabolic partial differential equation, we employ an implicit method of the Crank Nicholson variety which is well known in the literature \cite{Ames1992}.

Taking into account that the system has translational symmetry, i.e. \textbf{r} $\rightarrow$ z, the propagator is then only a function of $z$ and $s$. We denote $\partial_z^2 G$ as $G_{zz}$ and also $\partial_s \rightarrow G_s$ so that, upon discretizing:

\begin{subequations}
\begin{equation}
G_{zz}(s_{k+1},z_j) = \frac{G_{j+1}^{k+1}- 2G_{j}^{k+1} + G_{j-1}^{k+1}}{\Delta z^2}
\label{eq_sub_gzz}
\end{equation}
\begin{equation}
G_s (s_{k+1},z_j)= \frac{G_j^{k+1} - G_j^k}{\Delta s}
\label{eq_sub_gs}
\end{equation}
\end{subequations}

where $k$ and $j$ are discrete time and space labels respectively. Equation \ref{eq_FP} now becomes the discrete Fokker-Plank equation:

\begin{equation}
\frac{G_j^{k+1} - G_j^k}{\Delta s} = \frac{1}{6}(\frac{G_{j+1}^{k+1}- 2G_{j}^{k+1} + G_{j-1}^{k+1}}{\Delta z^2}) - w_j^{k+1} G_j^{k+1}
\label{eq_FPdis}
\end{equation}

which upon rearranging for $G_j^k$ we find: 

\begin{equation}
G_j^k = (1 + 2H + \Delta s w_j^{k+1})G_j^{k+1} - H(G_{j+1}^{k+1} + G_{j-1}^{k+1})
\label{eq_gjk}
\end{equation}

for which we have defined $H = \frac{\Delta s}{\Delta z^2 6}$ for simplicity. We need to solve this for $G$ at a later time $k+1$ given that $G^k_j$ is known, this entails solving a tridiagonal linear system of equations. For this process we use the following:

\begin{enumerate}
\item[$\bullet$] $G^k_j = G(z_j, s_k)$ where $z_j = j \Delta z$ for $(j=0, \cdots, n+1)$.
\item[$\bullet$] $\Delta z = \frac{L_z}{n+1}$, $s_k = k \Delta s$ for $(k=0,....,m)$ and $\Delta s =\frac{N}{m}$.
\item[$\bullet$] $N$ is the polymerisation of the chain.
\end{enumerate}

We can show this as a matrix equation (shown here for $j=0,1,2,3,4$):

\begin{equation}
\begin{bmatrix}
(1 + 2H + \Delta s w_1^{k+1}) & -s & 0 \\
-s & (1 + 2H + \Delta s w_2^{k+1}) & -s \\
0 & -s &(1 + 2H + \Delta s w_3^{k+1})
\end{bmatrix} \times 
\left(
\begin{array}{c}
G_1^{k+1} \\
G_2^{k+1} \\
G_3^{k+1}
\end{array}
\right) =
\left(
\begin{array}{c}
G_1^{k} \\
G_2^{k} \\
G_3^{k}
\end{array}
\right) +
H 
\left(
\begin{array}{c}
G_0^{k+1} = 0 \\
0 \\
G_4^{k+1} = 0
\end{array}
\right)
\label{eq_matrixnum}
\end{equation}

here the vector on the far RHS embodies the boundary conditions which in this case are of Dirichlet type. So finally solving for $G_j^{k+1}$ requires inverting the matrix on the LHS:

\begin{equation}
\boxed{\overline{G}^{k+1}_{j} = [C]^{-1}( \overline{G}^k_j + H \overline{bc})}
\end{equation}

where $[C]$ is the matrix as shown in equation \ref{eq_matrixnum} and bc embodies the boundary conditions. This equation is solved within a loop iterating time, so for each $s$ all the propagators are found which continues until the end of the chain.

The initial conditions are imposed at the first iteration of the time loop. So for the Dirac delta initial condition \ref{eq_i2} we have

\begin{lstlisting}[frame=single]
for( ... ) // loop over all space
{
	if(j*dz == dz)
		{G2 = 1*(N/A)*(1/G1);}
	else;
	
....
}
\end{lstlisting}

\newpage

\begin{figure}[h]
\begin{center}
\includegraphics[scale=0.5]{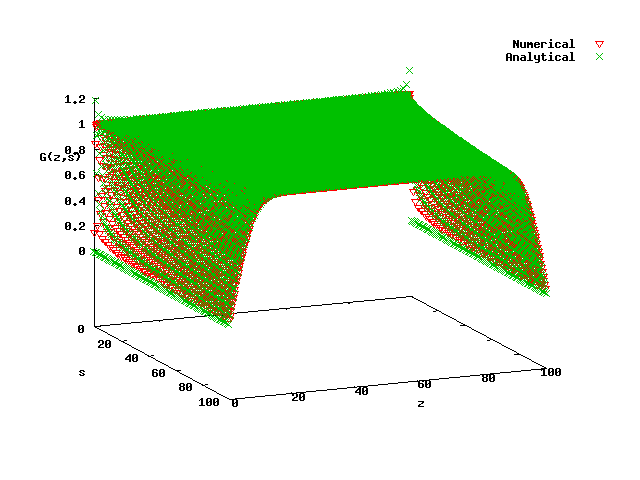}
\caption{ The greens function for an ideal chain of length 100 where $L_z = 100$ is compared to the analytical solution. The analytical function sums an infinitie series and here we computed 200 terms hence the spikes towards the boundary conditions. They otherwise match very well. One can now see the relation between this problem and finding the temperature distribution in a rod of length $L_z$.\label{fig_g1compare}}
\end{center}
\end{figure}

The propagator with the unity initial condition is compared to the analytical solution (for the ideal case) in figure \ref{fig_g1compare} to check that the numerical scheme works.

Note that the propagators are found for a particular form of the external field $w(z)$. If the whole system is to be solved self- consistently and equilibrium density profiles are to be found, we must employ another scheme which continually updates the mean-field with the segment density and vice versa. We use classical density functional theory to do this which will be outlined in the next section.

\section{Classical Density Functional Theory}

Density functional theory was originally formulated to describe the electronic structure of systems thermodynamically \cite{Kohn1996}. It was later reformulated to apply to classical systems of inhomogeneous fluids \cite{Evans1979}. The essence of a DFT is that the Hamiltonian  $\mathcal{H}$ of a system can be written as an explicit \textit{functional} of the density $\rho(\textbf{r})$. We know that a system is considered to be in thermodynamic equilibrium when the free energy $\mathcal{F}$ is minimized, so the general procedure of a DFT is to carefully 'design' a free energy functional and minimize this w.r.t to the classical density i.e. $\frac{\delta \mathcal{F}[\rho]}{\delta \rho} = 0$. A pedagogical introduction to cDFT is presented here \cite{Jeanmairet2014} as a computational experiment.

The word 'design' is used here because in general the form of $\mathcal{F}$ is not known a priori. Physical assumptions and considerations about the constituents of the system and how they interact with each other and the system boundaries are, essentially, plugged into the free energy. This is done through an excess term $\mathcal{F}_{ex}$ in addition to an ideal term $\mathcal{F}_{id}$ such that $\mathcal{F} = \mathcal{F}_{id} + \mathcal{F}_{ex}$. We now begin to 'design' a free energy functional for our polymeric system. In general the Hamiltonian is of the form:

\begin{equation}
\mathcal{H} = T + \frac{1}{2}  \sum_{j=1}^{N} \sum_{i \neq j}^N \phi( \textbf{r}_i - \textbf{r}_j)
\end{equation}

where $T$ is the kinetic term and $\phi$ is a pair potential, the factor of $\frac{1}{2}$ avoids double counting. 

We split the Hamiltonian into two parts $\mathcal{H} = \mathcal{H}_0 + \mathcal{H}_1$ where each term is defined by

\begin{subequations}
\begin{equation}
\mathcal{H}_0 = \sum_{i=1}^N \frac{\textbf{P}_i^2}{2 m_i} + \sum_{i=1}^N w(\textbf{r})
\end{equation}
\begin{equation}
\mathcal{H}_1 = \frac{1}{2} \sum_{j=1}^N \sum_{i \neq j}^N \phi(\textbf{r}_i - \textbf{r}_j) - \sum_{i=1}^N w(\textbf{r})
\end{equation}
\end{subequations}

where the mean field $w$ has units $\frac{1}{k_{B} T}$. We employ the Bogoliubov inequality $\mathcal{F} \leqslant \mathcal{F}_0 + <\mathcal{H}_1>_0$. Where the angular brackets means taking the ensemble average w.r.t to the ensemble of the ideal system. $\mathcal{F}_0 = -k_{B} T ln[Z_0]$ is the free energy of the ideal reference system. Putting things together using the preceding formulas we have

\begin{equation}
\mathcal{F} \leqslant -k_{B} T ln[Z_0] - <\sum_{i=1}^N w(\textbf{r}_i)>_0 + <\frac{1}{2} \sum_{j=1}^N \sum_{i \neq j}^N \phi(\textbf{r}_i - \textbf{r}_j)>_0
\label{eq_FEbra}
\end{equation}

where the $2^{nd}$ and $3^{rd}$ terms are replaced with $\int \rho(\textbf{r}) w(\textbf{r}) d\textbf{r}$ and $\int \rho(\textbf{r}) \rho(\textbf{r}^\prime) \phi(\textbf{r} - \textbf{r}^\prime) d\textbf{r} d\textbf{r}^\prime$ respectively.

We split the pair potential $\phi$ into an attractive and repulsive part. The repulsive part is embodied in an equation of state term known as the Carnahan - Starling equation \cite{Osmanovic2013} which imposes that the beads are hard spheres (HS) and are connected in a chain-like (CC) manner:

\begin{equation}
\begin{matrix}
\mathcal{F}^{CS} = & \int \rho(\textbf{r}) [ \frac{4 \eta (\textbf{r})  - 3\eta (\textbf{r})^2}{[1 - \eta(\textbf{r})]^2} - & (1 -\frac{1}{N}) ln (\frac{2 - \eta(\textbf{r})}{2[1-\eta(\textbf{r})]^3})] d \textbf{r} \\
& \downarrow & \downarrow \\
 & HS & CC
\end{matrix}
\label{eq_CS}
\end{equation}

here $\eta (\textbf{r}) = \frac{\pi d^3}{6} \rho(\textbf{r})$ (d $=$ 1nm) is the packing fraction of spheres in three dimensions. The attractive part of the potential between the beads is modelled as a simple exponential decay of the form:

\begin{equation}
u (\textbf{r}^\prime - \textbf{r}) = - \epsilon \times exp(-\frac{|\textbf{r}^\prime - \textbf{r}| - d}{\lambda})
\label{eq_att}
\end{equation}

for $|\textbf{r}^\prime - \textbf{r}| \geqslant d$ and where $\epsilon= $  the strength and $\lambda= $ the range of the interaction. 

\begin{figure}[h]
\begin{center}
\includegraphics[scale=0.4]{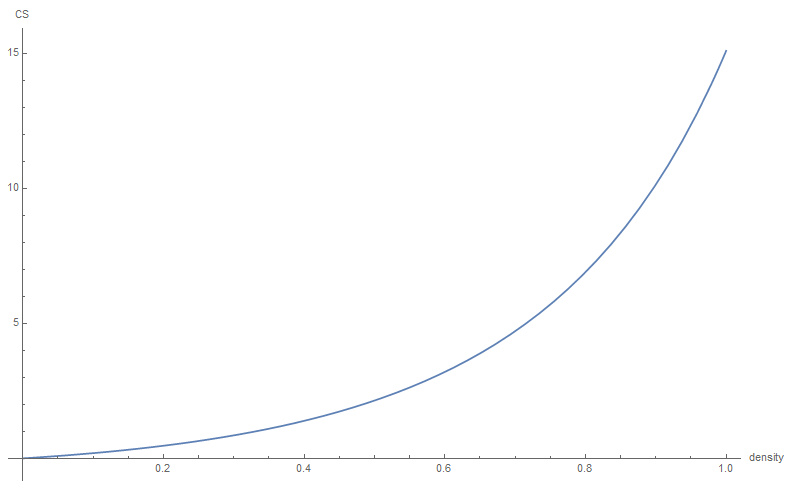}
\caption{How $\frac{\delta \mathcal{F}^{CS}[\rho]}{\delta \rho(\textbf{r})}$ behaves with increased density for $N=100$ monomers and for $d=1$. \label{fig_mathcomp}}
\end{center}
\end{figure}

We are now ready to explicitly state the mean field free energy of our system for $N_p$ polymers:

\begin{equation}
\boxed{\mathcal{F}_{mf} = - N_p ln[Z_0] + \mathcal{F}^{CS}[\rho(\textbf{r})] - \int \rho(\textbf{r}) w(\textbf{r}) d\textbf{r} + \frac{1}{2} \int \rho(\textbf{r}) \rho(\textbf{r}^\prime) u(\textbf{r} - \textbf{r}^\prime) d\textbf{r} d\textbf{r}^\prime}
\label{eq_Fmf}
\end{equation}

where the density is interpreted as a superposition of bead density profiles of $N_p$ polymers. This is because the problem, tackled in (SCFT), has essentially been reduced to a problem for one polymer in the presence of a mean field. In order to minimize the free energy and ascertain the equilibrium density profile one must continually update the mean field so it is no longer changing meaningfully. This means that the configuration of our system has essentially 'settled down', they are no longer rearranging themselves because they are comfortable in an equilibrium state. Precisely this means that:

\begin{equation}
\partial_t w(\textbf{r}, t) = \delta_\rho \mathcal{F}_{mf} = 0
\label{eq_std}
\end{equation}

where $t$ is a self invoked time variable which has no physical meaning, it represents a stage in an iterative scheme to continually update the mean field. Equation \ref{eq_std} is known as the steepest descent method. Discretizing this we can generate the next mean field by:

\begin{equation}
w^{k+1}(\textbf{r}) = w^{k} + \Delta t[-w^k (\textbf{r}) + (\frac{\delta \mathcal{F}^{CS}[\rho]}{\delta \rho(\textbf{r})})^k + \sum_j \rho^{k}(\textbf{r}^\prime) u(\textbf{r}^\prime - \textbf{r}) \Delta \textbf{r}]
\label{eq_stddis}
\end{equation}
\newpage

where the sum in the last term is over all space. The functional derivative of the HS and CC contributions in equation \ref{eq_stddis} should become larger for increasing density which would decrease the probability to go there. It is useful to check that this term implemented in code behaves and that the derivatives have been performed correctly, one could do this in Mathematica e.g. see figure \ref{fig_mathcomp}.

\begin{center}
\textbf{The Iterative Scheme}
\end{center}

\begin{enumerate}
\item Initially guess $w^0(z)$ is random in space.
\item Numerically solve for the propagator $G$ (unity IC) and then for $G_C$ (see equation \ref{eq_i2}).
\item Compute the density $\rho (z)$ using equation \ref{eq_density_muller}.
\item Update the mean field via equation \ref{eq_stddis} with $\Delta t$ such that the system converges (0.05 here).
\item Check \textbf{IF} $( w^{k+1} - w^k) \leq \gamma$ \footnote{$\gamma << 1$ but should be tailored for each simulation.}\textbf{DO} Finish simulation. \newline
\textbf{ELSE} Go to step 2. 
\item Plot results.
\end{enumerate}
\newpage
\section{Results}

\subsection{Testing the Mean Field}
\begin{figure}[h]
\begin{center}
\includegraphics[scale=0.4]{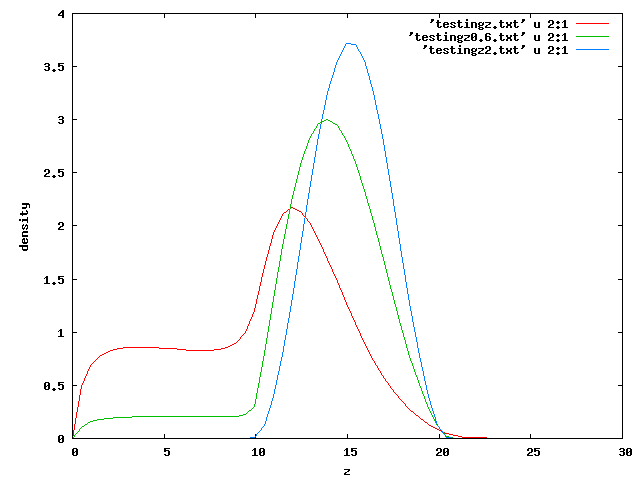}
\end{center}
\caption{ A square negative potential between $z=10 and z=20$ for increasing well depth. The well starts shallow (red) where there is some attraction, which is increased slightly (green) and at the greatest well depth (blue) the distribution becomes Gaussian. \label{fig_well}}
\end{figure}

To check that the mean field scheme is working it is recommended that one tests that the mean fields behaviour makes physical sense. A potential well with increasing depth was tested (figure \ref{fig_well}) to ensure that the monomers become attracted to it, which increases the density in that area. 

\begin{figure}[h]
\begin{center}
\includegraphics[scale=0.4]{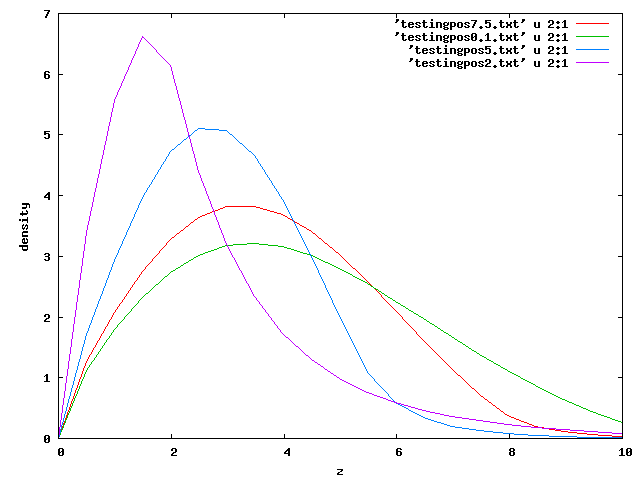}
\end{center}
\caption{ A positive mean field encroaches towards the grafting plane which leads to monomers squishing towards the impenetrable surface. \label{fig_poswell}}
\end{figure}

We also check that a positive 'hill' mean field leads to the density moving away from that region, which would correspond to an excluded volume interaction. This was tested by increasing the region for a constant mean field towards the grafting plane $z_\bot$ as shown in figure \ref{fig_poswell}.

\newpage
\subsection{Investigating Excluded Volume Effects}

\begin{figure}[h]
\begin{center}
\includegraphics[scale=0.5]{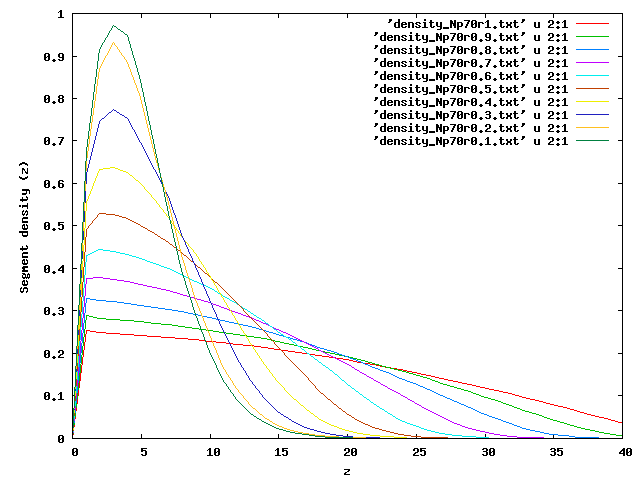}
\end{center}
\caption{ The diameter was decreased from the kuhn length in steps of 0.1. The greatest diameter (dark red) and the smallest diameter (dark green). \label{fig_results_d}}
\end{figure}

\begin{figure}[h]
\begin{center}
\includegraphics[scale=0.4]{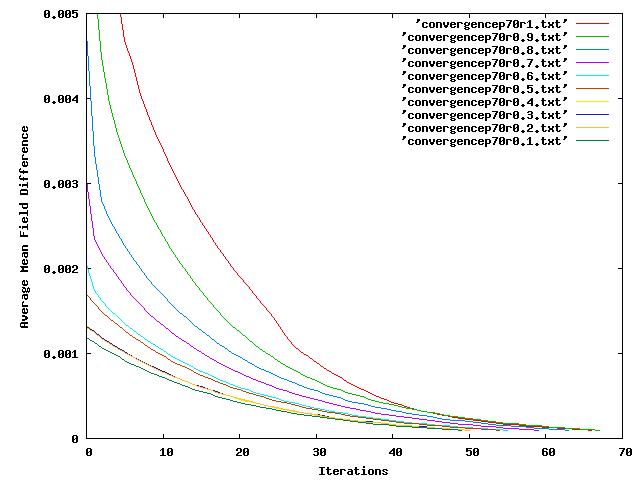}
\end{center}
\caption{ All simulations converged beyond the convergence criterion $\gamma = 0.0001$. \label{fig_results_dcon}}
\end{figure}

The iterative numerical cDFT scheme was used to investigate how the segment density behaves for different diameters ,$d$, of beads. The number of polymers used was 70 for $N=100$ monomers per polymer and the area, $L_\bot$, was chosen such that grafting density is high enough that polymers interact with one another. For the numerical solution of the propagators $\Delta s = 1$ and the spatial grid spacing was tuned as to gain sufficient resolution to keep the code as fast as possible. Figure \ref{fig_results_d} shows plots for increased bead diameter and figure \ref{fig_results_dcon} shows the convergence of the mean field for each simulation run.

\begin{figure}[h]
\begin{center}
\includegraphics[scale=0.45]{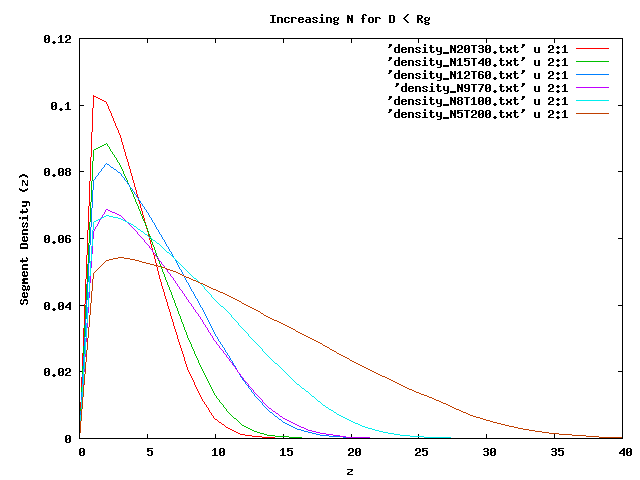}
\end{center}
\caption{ Plots for $N$ is 200 (brown), 100 (light blue), 70 (purple), 60 (blue), 40 (green) and 30 (red). For all plots $d = 1$.  \label{fig_results_N}}
\end{figure}

The polymerisation $N$ was changed to investigate how it would effect the formation of polymer layers shown in figure \ref{fig_results_N}, the grafting density was shifted for each $N$ to satisfy $D < R_F$. 

\begin{figure}[h]
\begin{center}
\includegraphics[scale=0.45]{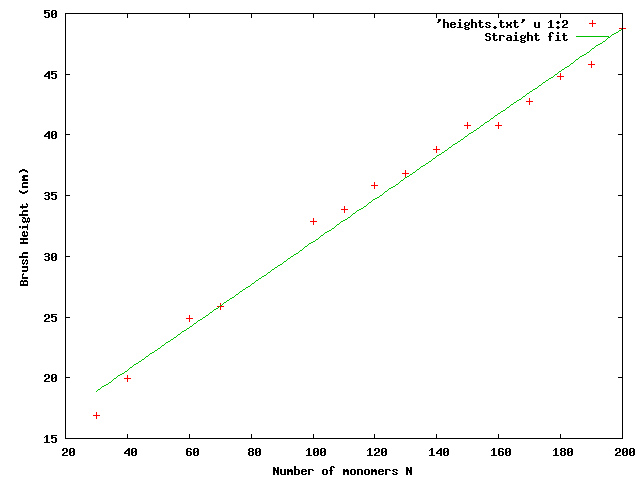}
\end{center}
\caption{ Heights computed for polymerisation $N$ and a straight line fitted using the Gnuplot software. \label{fig_results_height}}
\end{figure}

The brush height $h$, which is defined as the maximum $z$ value of the system of polymers, was plotted against polymerisation $N$ which is shown in figure \ref{fig_results_height}.

\newpage 

\subsection{Discussion}

By either increasing the diameter of the beads or the number of beads per chain one reduces the amount of available space for the beads to move around in. In the presence of excluded volume interactions the polymer chains prefer to extend outwards from the grafting plane which is thermodynamically preferable. 

This extending outwards of the chain produces a polymer brush and reducing the excluded volume interactions either by decreasing $d$, increasing $D$ or reducing $N$ leads to the mushroom layer as first shown in figure \ref{fig_mush}. Where the monomers are more densely packed towards the grafting plane. 

The brush height $h$ scales linearly with $N$, for $N \geq 60$, which agrees with the scaling argument of Alexander \cite{Jones1999} where $h \approx (\sigma)^{1/3} N $.

Another interesting point of discussion is how the parabolic profile of Milner (figure \ref{fig_MWC}) differs to the results shown here. Already taking into account that the boundary conditions at the surface are implemented differently one must also keep in mind that the SCFT of MWC is for very long chain lengths. Hence at $h_{max}$ the profiles should differ slightly, in fact they should be less abrupt than that of MWC. 

\newpage
\bibliography{library}
\bibliographystyle{unsrt}
\end{document}